\documentclass[twocolumn,superscriptaddress,aps,showpacs,floatfix,prl,10pt]{revtex4-1}
\usepackage[T1]{fontenc}
\usepackage{amssymb,amsmath,bm}
\usepackage{graphicx}
\usepackage{color}
\usepackage{bbold}
\usepackage{bbm}
\usepackage{appendix}
\usepackage{soul}
\usepackage[dvipsnames]{xcolor}
\usepackage{ulem}
\usepackage{float}
\usepackage{textcomp}
\usepackage{ mathrsfs }
\usepackage{siunitx}

\newcommand{\RNum}[1]{\uppercase\expandafter{\romannumeral #1\relax}}

\begin{document}
\title{Superpolynomial Quantum Enhancement in Polaritonic Neuromorphic Computing}
\author{Huawen Xu}
\email{huawen001@e.ntu.edu.sg}
\affiliation{Division of Physics and Applied Physics, School of Physical and Mathematical Sciences, Nanyang Technological University, 21 Nanyang Link, Singapore 637371, Singapore}

\author{Tanjung Krisnanda}
\affiliation{Division of Physics and Applied Physics, School of Physical and Mathematical Sciences, Nanyang Technological University, 21 Nanyang Link, Singapore 637371, Singapore}
\author{Wouter Verstraelen}
\affiliation{Division of Physics and Applied Physics, School of Physical and Mathematical Sciences, Nanyang Technological University, 21 Nanyang Link, Singapore 637371, Singapore}
\author{Timothy C. H. Liew}
\email{timothyliew@ntu.edu.sg}
\affiliation{Division of Physics and Applied Physics, School of Physical and Mathematical Sciences, Nanyang Technological University, 21 Nanyang Link, Singapore 637371, Singapore}
\affiliation{MajuLab, International Joint Research Unit UMI 3654, CNRS, Universit\'e C\^ote d'Azur, Sorbonne Universit\'e, National University of Singapore, Nanyang Technological University, Singapore}

\author{Sanjib Ghosh}
\email{sanjib.ghosh@ntu.edu.sg}
\affiliation{Division of Physics and Applied Physics, School of Physical and Mathematical Sciences, Nanyang Technological University, 21 Nanyang Link, Singapore 637371, Singapore}
\begin{abstract}
Recent proof-of-principle experiments have demonstrated the implementation of neuromorphic computing using exciton-polaritons, making use of coherent classical states [D.~Ballarini \emph{et al}., Nano Lett. {\bf20}, 3506 (2020)]. At the same time, it is expected that nonlinear exciton-polaritons can reach a quantum regime forming non-classical states. Here we consider theoretically the quantum nature of exciton polaritons and predict a superpolynomial quantum enhancement in image recognition tasks. This is achieved within experimentally accessible parameters.
\end{abstract}
\maketitle

\section{\RNum{1}. INTRODUCTION}The Exciton-polariton has emerged as an attractive quasiparticle, a hybrid of light (photon) and matter (exciton) in semiconductor microcavities, which possesses the characteristics of long dephasing time and ultrafast response {inherited} from their photonic part as well as strong nonlinearity from their excitonic part~\cite{Carusotto2013,Kasprzak2006}. 
These properties have encouraged works developing polaritonic information processing devices~\cite{Sanvitto2016,Fraser2017}. 
In particular, the experimental realization of various processing components has been reported,  including: logic gates~\cite{Leyder2007,Cancellieri2015}, diodes~\cite{Nguyen2013}, amplifiers~\cite{Niemietz2016}, switches~\cite{Gao2012,De-Giorgi2012,Dreismann2016,Grosso2014}, couplers~\cite{Rozas2020,Klaas2019}, junctions~\cite{Winkler2017}, transistors~\cite{Ballarini2013,Lewandowski2017a,Zasedatelev2019}, routers~\cite{Flayac2013,Marsault2015}, high speed transmission lines~\cite{Lerario2017} and memories~\cite{Cerna2013}. Although individual elements are well studied with exciton-polaritons, very few experimental works have studied their combination into complete information processing systems.

Theoretical works have considered that rather than imitating CMOS logic, exciton-polaritons might be more suited to alternative architectures of information processing, such as simulators of other systems~\cite{Kalinin2019} or Hamiltonian optimizers~\cite{Kalinin2018}, cellular automata~\cite{Banerjee2020}, or neural networks~\cite{Espinosa-Ortega2015}. These proposals have been based on the high degree of tuneability available in exciton-polariton systems, which is particularly useful in neural networks where functionality is determined by the strength of weight connections between network nodes. In general, neural network architectures serve as a good platform to apply machine learning algorithms, which circumvent the need to encode the function of the network in its design directly but rather allow it to be learned. Indeed, simple feedback algorithms have been used in exciton-polariton experiments to induce specific patterns~\cite{Ohadi2017, 2020arXiv200706690T}.
However, so far the implementation of the most well-known feedforward neural networks has not been attempted with exciton-polaritons, partly due to the large number of weight connections that would need individual control.

The same challenge of needing to simultaneously control many weight connections has been avoided in many other hardware realizations of neural networks by using so-called reservoir computing networks~\cite{Tanaka2019}. These architectures are based on completely randomly connected distributions of network nodes~\cite{Jaeger2004,Lukosevicius2009,Enel2016}, where all training is implemented in a single output layer, which also greatly simplifies learning algorithms~\cite{Lukosevicius2012}. Reservoir computing was considered theoretically for exciton-polariton systems \cite{Opala2019,Xu2020} and implemented experimentally shortly after~\cite{Ballarini2020,Mirek2021}, where benchmarking for the task of recognizing a standard set of hand-written digits showed the exciton-polariton system to have a higher success rate than previous hardware implementations of reservoir computing.

All the aforementioned works on exciton-polaritons have operated with classical coherent states, formed either through direct laser excitation or the formation of a polariton condensate. {On the other hand, polaritons are quantum particles, which has been evidenced in different experiments. In particular, the onset of non-classical correlations~\cite{Savasta2005} and squeezing~\cite{Karr2004,Boulier2014} has been reported, while more recent experiments showed the preservation of polariton entanglement~\cite{Cuevas2018} and imaging of the single polariton wavefunction~\cite{suarez2020quantum}. Other recent experiments~\cite{munoz2019emergence,delteil2019towards} have shown the onset of the polariton blockade~\cite{Verger2006}, where antibunching is generated from the interactions between polaritons themselves. Although the amount of antibunching measured was limited by the weak polariton-polariton interaction strength, a number of methods exist for increasing it~\cite{Kyriienko2020,Rosenberg2018,Jia2018,Stefanatos2020}.

Here, we consider the performance of polariton reservoir computers accounting for their quantum nature. In principle, the larger Fock space can be expected to allow a quantum enhancement, as has been demonstrated for quantum reservoirs based on spin systems~\cite{Fujii2017,Nakajima2019,Chen2019,Chen2020} and a single nonlinear oscillator~\cite{Govia2021} or a qudit~\cite{2021arXiv210111729K}. As polariton reservoir computers have previously been benchmarked for image recognition, we will consider images with their pixels converted into a time-dependent signal that excites the quantum polariton reservoir with a resonant amplitude and/or phase modulated laser.
The output from the quantum polariton reservoir is represented by the quasiprobability distribution in phase space, namely its Wigner function \cite{Olivares2012}, which will be used to classify the input image.
By comparing to simulations with classical polariton reservoirs, we find that the quantum polariton system is able to classify images with a single polariton node! We will continue to refer to this node as the ``reservoir'' as it nevertheless has a significant Fock space and fulfils the role of the much larger reservoir of classical reservoir computers, which cannot attain the same error rate with even hundreds of classical nodes. The single quantum polariton node exceeds also the limit of a classical linear classifier and is found empirically to have a superpolynomial quantum enhancement.  We also compare possibilities using two-photon excitation, which allows further enhancement.

\section{\RNum{2}. Quantum polariton reservoir}The physical model considered here consists of only a single polariton mode, {where} the Fock states of the quantum mode form the quantum reservoir network, {see Fig.~\ref{Figure1}$\textbf{a}$.} 
The bare Hamiltonian $ \mathcal{\hat H}_0$ for the polariton mode {is taken as}
\begin{eqnarray}
\mathcal{\hat H}_0= E\hat a^ \dagger \hat a + \alpha \hat a^ \dagger \hat a^ \dagger \hat a \hat a, 
\label{Equation1}
\end{eqnarray}
where $\hat a^ \dagger $ {($\hat a$)} is the creation (annihilation) operator, $E$ is the energy of the polariton mode and $\alpha$ is the Kerr nonlinear interaction strength. {For encoding information, we introduce a coherent single photon ($\mathcal{SP}$) and two photon ($\mathcal{TP}$) pumping, such that the total Hamiltonian (in a rotating frame at the frequency of the pump $\omega_p$) is written as
\begin{eqnarray}\label{Equation2}
\mathcal{\hat H} &=& \Delta \hat a^ \dagger \hat a + \alpha \hat a^ \dagger \hat a^ \dagger \hat a \hat a \nonumber\\
&&+ F(\hat a^{\dagger}  +  \hat a)+ P( e^{i\Theta}\hat a^{\dagger} \hat a^{\dagger} + e^{-i\Theta}  \hat a  \hat a), 
\end{eqnarray}
where $F$ and $P$ are real amplitudes of the $\mathcal{SP}$ and $\mathcal{TP}$ pumping, respectively, and we allowed for a phase difference of the $\mathcal{TP}$ pump, denoted by $\Theta$.
The detuning term is defined as $\Delta=E-\hbar \omega_p$, and we have assumed that the pumping frequency of $\mathcal{SP}$ ($\mathcal{TP}$) is $\omega_p$ ($2\omega_p$).
See also Ref.~\cite{bartolo2016exact}, in which a similar treatment is presented. 
The network structure composed of the Fock states can be illustrated as follows.
The single pumping term in Eq.~(\ref{Equation2}) expressed in the Fock basis reads $\sum_{n=0} F\sqrt{n+1} (|n+1\rangle \langle n| +|n\rangle \langle n+1|)$. 
This way the $\mathcal{SP}$ is seen as nearest neighbour connection between the Fock states.
Similarly, the two photon pumping serves as the connection between next nearest neighbours.
In our analysis, we compare four different pumping scenarios: $\mathcal{SP}$ ($P=0$); $\mathcal{TP}$ ($F,\Theta=0$); $\mathcal{SPTP}$ ($\Theta=0$); as well as $\mathcal{SPTP}^{\prime}$ where $\Theta=\Theta(t)\neq0$.

\begin{figure}[b]
\includegraphics[width=1\columnwidth]{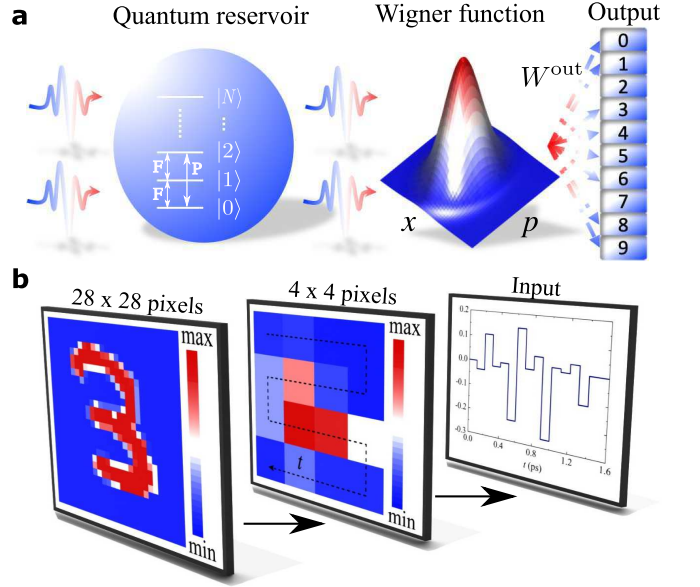}
\caption{{The schematic for image classification using quantum polariton reservoir computing. 
\textbf{a}) Input signals excite a single nonlinear polariton mode in the form of pumping amplitudes and/or phases. 
After a certain time $\tau$, the quasiprobability distribution in the phase space (Wigner function) is calculated.
By applying trained output weights ($\bm{W}^{\text{out}}$) to the Wigner function, a final output classifies the input signals (images).}
\textbf{b}) Each image (from the MNIST standard data set) is first compressed from $28\times28$ resolution to $4\times4$. 
The pixels of the compressed image {serve as time-dependent input signals.}}
\label{Figure1}
\end{figure}

In our simulations, we evolve the density matrix $\rho$ under {the} different pumping methods, where the initial state of the polariton mode is given by the vacuum state.
The dynamics of the system can be described by the quantum master equation:
\begin{eqnarray}
i\hbar {\dot \rho}= [\mathcal{\hat H}, \rho] +\frac{i\gamma}{2} \mathscr{L}(\rho,\hat a).
\label{Equation3}
\end{eqnarray}
where $\mathcal{\hat H}$ is the Hamiltonian of Eq.~(\ref{Equation2}), $\mathscr{L}(\rho,\hat a)$ is the Lindblad dissipation superoperator defined as $\mathscr{L}(\rho,\hat a)  \equiv 2 \hat a \rho \hat a^{\dagger} - \hat a^{\dagger} \hat a \rho - \rho \hat a^{\dagger} \hat a $, and $\gamma$ is the dissipation rate. 
As can be seen from Fig.~\ref{Figure1}$\textbf{a}$, the input information is first fed into the single polariton node.
After a finite time $\tau$ of evolution, {the Wigner function of the system is calculated from the density matrix $\rho_{\tau}$.
We note that for experiments, there are various well-developed methods for measuring Wigner function. 
For example, it can be reconstructed from the homodyne measurements of the rotated quadrature phase~\cite{vogel1989determination} or direct measurements based on photon counting~\cite{banaszek1999direct}.
In the present scheme, this may be equivalently performed on the emitted light from the microcavity. 
The Wigner function is then sampled discretely on a grid in phase space and the samples arranged into a vector, $\mathbf{f}$.
A final signal is calculated as $\bm{Y}^{\text{final}} = \bm{W}^{\text{out}} \bm{X}$, which is capable of input classification.
Here $\bm{W}^{\text{out}}$ is a trained output weight matrix and the vector $\bm{X}=[1,\textbf{u},\textbf{f}]$, which is allowed to also contain the input data pixels $u_i$ ($i=1,2,\cdots,16$) obtained as follows.}
The input image {with $28\times28$ resolution is compressed into $4\times4$ pixels (an equivalent procedure has been used in Ref. \cite{Ballarini2020}), which are then} converted into time-dependent signals with alternating signs, such that each pixel corresponds to a different {input signal value $u_i$} in time, {as shown in Fig.~\ref{Figure1}$\textbf{b}$}. 
{Note that even though the reduced $4\times 4$ images are rather unrecognizable to humans, they can still be recognized with the machine trained to do so.
We show below that our method can classify the $4\times4$ resolution images and perform better than classical polariton computers.}
{Note that all the input signal values $\textbf{u}$ are weighted with a random mask (a $16\times16$ matrix $\bm{W}^{\text{in}}$), where its elements are randomly sampled from $0$ to ${W}^{\text{in}}_{\text{max}}$, before this image information enters the quantum reservoir as time-dependent pumping amplitudes ($F$ and $P$) or phase ($\Theta$).}

\section{\RNum{3}. Performance}
Note that we have used $\Delta/\gamma=0.1$, $\alpha/\gamma=0.1$ and $\gamma \tau /\hbar=1.2$ in our simulations.
The $W^{\text{in}}_{\text{max}}$ is optimised for pumping amplitudes $F/\gamma$, $P/\gamma$ and the phase $\Theta$.
The training procedure is performed on the output weights $\bm{W}^{\text{out}}$ with ridge regression using a set of randomly chosen and known images from the MNIST database.
After the training, we keep the quantum reservoir fixed and use a set of unknown images to test and evaluate its efficiency for the image classification task.
Details of the supervised training of the polariton reservoir computing can be found in Appendix A.

We fix the number of training samples at $4000$ and testing samples at $1000$.
A separate analysis using a linear classifier method \cite{Lecun1998}, results in a classification error rate of $29.8\%$ and $26.4\%$ for normalized and unnormalized input signals $\textbf{u}$, respectively.
Due to the compression of the resolution of each image, it is difficult to achieve a low error rate with a linear classifier.
For our quantum polariton computing, however, considerably lower error rate is possible.
Our results for four different input encoding methods {($\mathcal{SP},\mathcal{TP},\mathcal{SPTP},$ and $\mathcal{SPTP}^{\prime}$)} are shown in Fig.~\ref{Figure2}. Details of the influence of physical parameters in the system are presented in Appendix B. To assess the scaling of the performance, rather than considering scaling with the system size, we consider scaling with a measure of the number of Fock states available to the quantum reservoir. To do this, we introduce an effective cutoff in the Fock space size, $N$. This effective cutoff $N$ naturally appears for a finite pumping strength, which can only inject a finite number of polaritons in the mode. Of course, in a physical system the Fock space is unbounded {(which can be accessed by increasing the pumping strength)}, but we wish to emphasize that {the scaling of Fock space size (which grows quadratically with $N$) can be enhanced by the appropriate choice of information encoding.} For different pumping methods, the performance is {indeed} found to improve from $\mathcal{SP}$ to {$\mathcal{TP}$ to $\mathcal{SPTP}$}, and finally to $\mathcal{SPTP}^{\prime}$. 
This improvement originates from the different ways the Wigner function contains the information (distribution and squeezing) in regards to the different pumping scenarios, see details of the analysis in Appendix C.

\begin{figure}[h]
\includegraphics[width=1\columnwidth]{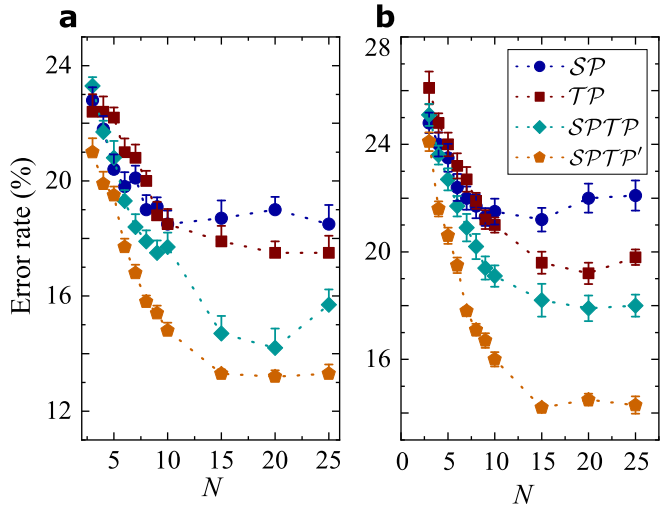}
\caption{Classification error rate for different quantum reservoir sizes (Fock space dimension) $N$ and different pumping methods: {labelled $\mathcal{SP}$ (single-photon); $\mathcal{TP}$ (two-photon); $\mathcal{SPTP}$ (single- and two-photon); and $\mathcal{SPTP}^{\prime}$ (single- and phase-dependent two-photon) as written in the Hamiltonian of Eq.~(\ref{Equation2}).}
Each data point is obtained {from the results of} $10$ random realizations of the input weights $\bm{W}^{\text{in}}$. 
Panels \textbf{a} and \textbf{b} present the lowest and average error rate among the 10 realizations of the input weights, respectively, of all the random realizations.}
\label{Figure2}
\end{figure}

\section{\RNum{4}. Classical polariton reservoir}To demonstrate the impressive performance of the quantum polariton reservoir in image recognition, we compare our system with classical polariton reservoir computing. 
In this case, we consider a square polariton lattice with $N_c$ polariton modes having random nearest-neighbour coupling in a tight-binding model.
The dynamics is described by the general discrete nonlinear Schr\"{o}dinger equation:
\begin{eqnarray}
i\hbar \frac{\partial \psi_n}{\partial t}&=&i \sum_{j=1,2}\bm{W}_{nj}^{\text{in}} \mathbf{v}_j +\sum\limits_{m\in \tilde n} \bm{K}_{nm} \psi_m + i\gamma \psi_n \nonumber \\
&&+ \alpha |\psi_n |^2 \psi_n , 
\label{Equation4}
\end{eqnarray}
where $\bm{W}^{\text{in}}$ (now a $N_c\times 2$ matrix) is an input mask having random weights $\bm{W}^{\text{in}}_{nj}\in[0,{W}^{\text{in}}_{\text{max}}]$, the vector $\mathbf{v}=[1,u]$ containing the input signal, $\bm{K}_{nm}$ the random nearest-neighbour coupling in the lattice, $\gamma$ the overall gain-loss profile, $\alpha$ the Kerr nonlinear interaction strength {and $\tilde n$ denotes the nearest neighbors of mode $n$.
Note that in the classical polariton reservoir computing, the obtained output from the reservoir is the intensity of each mode, i.e., $|\psi_n|^2$, whereas in the quantum polariton computing, we use the Wigner function.
Similarly to the quantum case, here $\bm{X}=[1,\textbf{u},\textbf{f}]$, where now the vector $\textbf{f}=[|\psi_1|^2,|\psi_2|^2,\cdots,|\psi_{N_c}|^2]$.

In simulations, we keep the physical parameters the same as used in the quantum scenario, i.e., $\alpha/\gamma = 0.1$. As illustrated in Fig.~\ref{Figure3}, by increasing the lattice size of the classical polariton reservoir, {the} error rate saturates at $\sim14\%$. To achieve the same (best) performance as in the quantum polariton reservoir computing, one requires a classical reservoir size of $100$ polariton modes for $\mathcal{SP}$ and $\mathcal{TP}$, and more than $250$ polariton modes for $\mathcal{SPTP}$. For the lowest error rate ($13\%$) obtained from $\mathcal{SPTP}^{\prime}$, the classical polariton computer still cannot be on par even with nearly $700$ polariton modes.
We note that the recent experiments in Ref.~\cite{Ballarini2020} reported $86\%$ success rate for $4\times 4$ resolution with a 400-node classical polariton reservoir.
In our case, this is achieved theoretically with a single quantum polariton mode.

\begin{figure}[h]
\includegraphics[width=0.95\columnwidth]{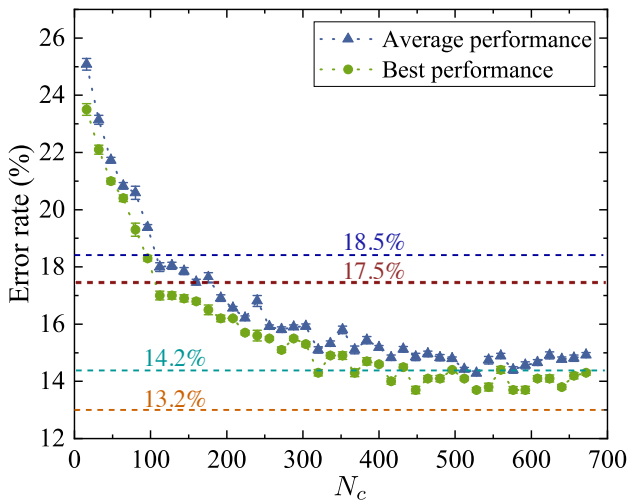}
\caption{Classification error rate (best and average performance) of the classical polariton reservoir {for} different reservoir sizes $N_c$. 
Each data point here is obtained {from} $10$ random realizations of the input weight matrix $\bm{W}^{\text{in}}$. 
The four dashed lines correspond to the best performance of the quantum polariton reservoir in Fig. \ref{Figure2}.}
\label{Figure3}
\end{figure}

\section{\RNum{5}. Quantum enhancement}
By comparison with classical polariton reservoir computing, the quantum polariton reservoir has a clear improvement in terms of performance and system size. {This is possible due to the quantum enhancement originating from the high dimensionality of the polariton Fock space. As a measure of quantum enhancement, the physical size of a quantum system is traditionally compared with required size of a classical system to achieve the same level of performance. In this respect, a classical reservoir computer requires hundreds of polariton modes to achieve the same level of success rates shown by only a single quantum polariton mode. This is clearly a superpolynomial quantum enhancement. However, }noting that a quantum polariton mode involves many Fock states, another comparison would be between the physical size of a classical reservoir and the effective number of qubits needed if one were to construct the Fock space of the quantum mode. This comparison is typically used to demonstrate quantum advantage in terms of information storage, see, e.g., Refs.~\cite{Tacchino2019,Markovic2020}.

The advantage in the quantum regime is illustrated in Fig.~\ref{Figure4}. 
The horizontal axes in Fig. \ref{Figure4} are defined as the effective number of qubits $N_{q}= \log_2N$, where $N$ is the effective Fock space dimension (minimum number of Fock states required to describe the system). Ideally, the Fock space dimension of a polariton mode is infinite. However, with a finite amount of pumping strength only a finite number of Fock states will be occupied, which defines a finite effective dimension.
 Each data point in Fig.~\ref{Figure4} corresponds to a specific error rate for image recognition; for example, the point indicated by the arrow in panel $\mathbf{b}$ represents achieving an error rate $\leq$ $14\%$, in which case it requires less than four effective qubits for the quantum polariton reservoir and $500$ modes for the classical polariton scenario.
It is apparent from Fig.~\ref{Figure4} that the quantum polariton reservoir showcases superpolynomial enhanced performance compared to the classical polariton reservoir.
Note that the number of classical modes $N_c$ is plotted in logarithmic scale in both panels.

\begin{figure}[h]
\includegraphics[width=0.95\columnwidth]{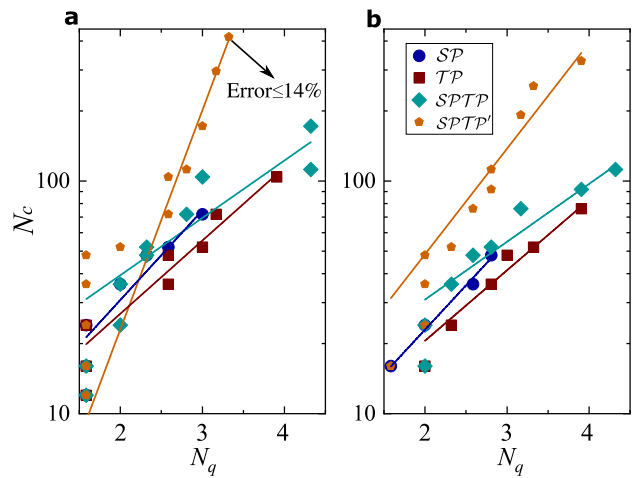}
\caption{Superpolynomial enhancement of the quantum polariton reservoir computing. 
The vertical axis in each panel represents the number of polariton modes required for achieving a specific error rate, while the horizontal axis is the number of effective qubits required in the quantum reservoir computing. 
Panels \textbf{a} and \textbf{b} represent the best and average performance, respectively, of the quantum and classical reservoir.}
\label{Figure4}
\end{figure}

\section{\RNum{6}. Conclusion} In this paper, a quantum reservoir network that is formed by the Fock space of quantum polaritons  is shown to possess the capability for effective image classification task. 
By utilizing different pumping methods, i.e., single-photon and phase-dependent two-photon pumping, we show performances with an error rate {as low as} $13\%$ with practical parameters (e.g., $\alpha / \gamma = 0.1$). 
Our results show that by exploiting the Fock space dimension of one quantum polariton mode, one can achieve lower error rate for image classification task, than that obtained from a classical reservoir with $\sim700$ polariton modes. 
Finally, by comparing the required system sizes of the quantum polariton reservoir (effective number of qubits) and the classical polariton case for achieving the same classification efficiency, superpolynomial quantum enhancement is demonstrated.
Our results motivate applications exploiting the quantum nature of exciton-polaritons.

\appendix
\renewcommand{\thefigure}{A\arabic{figure}}
\renewcommand{\theequation}{A\arabic{equation}}
\setcounter{equation}{0}
\section{APPENDIX A: Supervised Training}
In the supervised training, we offer the reservoir network a set of known images from the MNIST database, which correspond to the input signals $\mathbf{u}_{k}$ with the subscript denoting the $k$th image, where each $\mathbf{u}_{k}=[u_1,u_2,\cdots,u_{16}]$ and $k=1,2,\cdots,N_{tr}$. 
Also, before the signals enter the reservoir, we apply a random mask $\bm{W}^{\text{in}}$ to each of the input signals $\mathbf{u}_k$. 
With the known images, we have the corresponding desired target outputs $\mathbf{Y}^{\text{target}}_k$.
The output we obtain directly from the reservoir network is its Wigner function, arranged in a vector form $\mathbf{f}_k$. 
We take the output layer of the reservoir as a vector $\mathbf{X}_k = [1,\mathbf{u}_k,\mathbf{f}_k]$ and the relation between the final output and the output layer is given by $\mathbf{Y}^{\text{final}}_k= \bm{W}^{\text{out}} \mathbf{X}_k$. 
For a set of known examples, we have
\begin{eqnarray}
\mathcal{Y}^{\text{final}} = \bm{W}^{\text{out}} \mathcal{X}, 
\label{Equation5}
\end{eqnarray}
where $\mathcal{Y}^{\text{final}} = [\mathbf{Y}_1^{\text{final}};\mathbf{Y}_2^{\text{final}};...;\mathbf{Y}_{N_{tr}}^{\text{final}}]$ is the collection of all the final outputs and $\mathcal{X} = [\mathbf{X}_1;\mathbf{X}_2;...;\mathbf{X}_{N_{tr}}]$ is the corresponding collection of the obtained outputs. 
The purpose of the training is to minimize the difference between $\mathcal{Y}^{\text{final}}$ and $\mathcal{Y}^{\text{target}}$, which stands for the collection of all the target outputs, i.e., $\mathcal{Y}^{\text{target}}=[\mathbf{Y}_1^{\text{target}};\mathbf{Y}_2^{\text{target}};...;\mathbf{Y}_{N_{tr}}^{\text{target}}]$. 
We apply the ridge regression \cite{Hoerl1970} here to get the optimized output weight matrix $ \bm{W}^{\text{out}} $:
\begin{eqnarray}
\bm{W}^\text{out}= \mathcal{Y}^{\text{target}} {\mathcal{X}}^{T}( \mathcal{X} \mathcal{X}^{T}+{\beta}\bm{I})^{-1},
\label{Equation6}
\end{eqnarray}
where $\mathcal{X}^{T}$ is the transpose of the matrix $\mathcal{X}$, $I$ is an identity matrix and $\beta$ is the regularization coefficient. 
The term $\beta I$ is used to control the $\bm{W}^{\text{out}}$, in case there is small fluctuation in the input signals, the results still converge.

\renewcommand{\thefigure}{B\arabic{figure}}
\section{APPENDIX B: Analysis of Physical Parameters}

Before the input signals ($\mathbf{u}$) enter the reservoir network, a random mask $\bm{W}^{\text{in}}$ is applied to it. 
The elements in the mask are randomly distributed from $0$ to ${W}_{\text{max}}^{\text{in}}$.
This way, the actual input signals entering the reservoir will be $\bm{W}^{\text{in}} \mathbf{u}$. 
In other words, the amplitude of the input mask determines the pumping strength and/or phase amplitude. 
We take the $\mathcal{SP}$ pumping method as an example, vary the ${W}^{\text{in}}_{\text{max}}$ and analyse the classification error rate. 
As shown in Fig.~\ref{FigureB1}, with increasing of ${W}_{\text{max}}^{\text{in}}$, the classification error rate goes down.
\begin{figure}[h]
\setcounter{figure}{0}
\includegraphics[width=0.95\columnwidth]{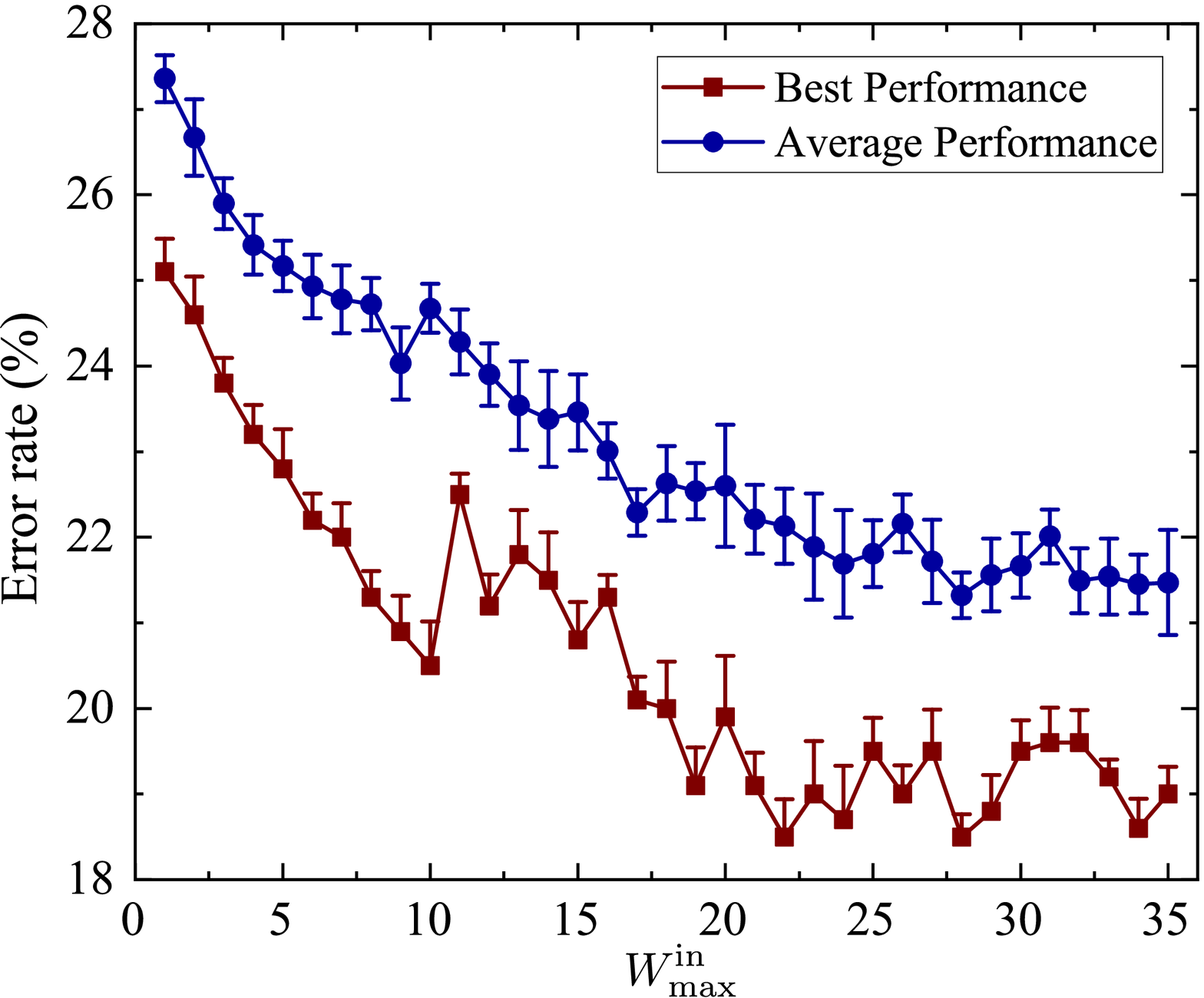}
\caption{Average and lowest classification error rate for different input weight amplitudes (${W}^{\text{in}}_{\text{max}}$), for the $\mathcal{SP}$ pumping scheme. Each point is obtained from $10$ random realizations of the matrix $\bm{W}^{\text{in}}$. $N$ is fixed at 10.}
\label{FigureB1}
\end{figure}

The significance of nonlinearity in the system is analysed as well for the $\mathcal{SP}$ pumping method. To make the role of nonlinearity more prominent, we vary $\alpha$ while fixing the randomly chosen mask $\bm{W}^{\text{in}}$ and $\gamma$. As can be seen in Fig.~\ref{FigureB2}, with increasing the nonlinearity, the classification error rate drops, which indicates the quantum polariton reservoir works better in stronger nonlinear regime.
\begin{figure}[h]
\includegraphics[width=0.95\columnwidth]{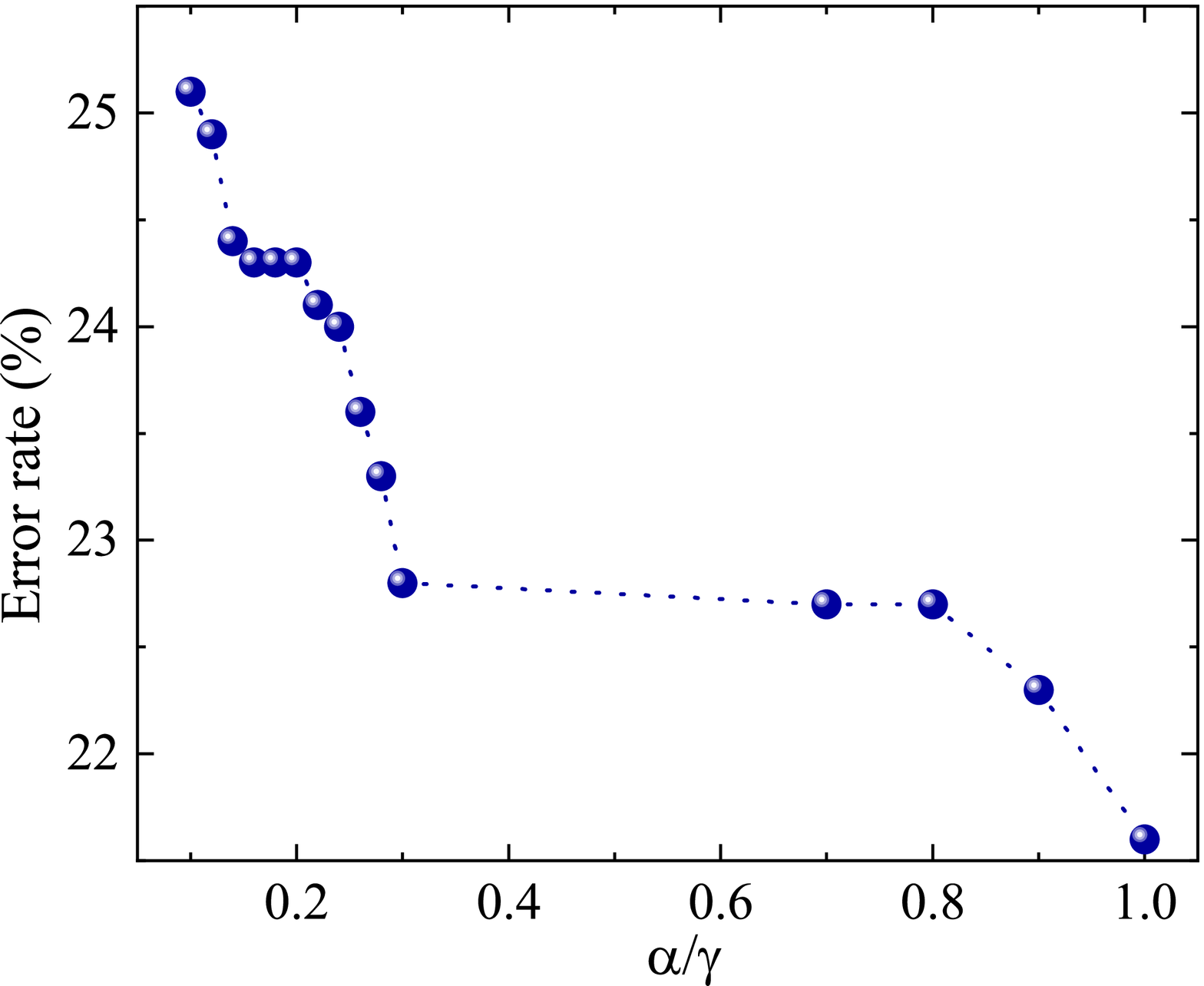}
\caption{Effect of nonlinearity in the system, showing the classification error rate under different nonlinear strength.}
\label{FigureB2}
\end{figure}

In real situations, there might be a fluctuation in the detuning $\Delta$ coming from the fluctuation in the driving laser frequency $\omega_p$. 
To investigate this, we added a random time-dependent energy fluctuation term $\eta(t)\in[-1,1]\delta$ (where $\delta$ is the fluctuation strength) in the detuning. 
As an example, we plotted the error rate against the fluctuation strength for the single photon pumping case ($\mathcal{SP}$) in Fig.~\ref{FigureB3}. 
It can be seen that the classification error rate is almost unchanged ($\sim24.5\%$), which holds true even if the fluctuation-to-linewidth ratio is $\delta/\gamma \sim0.1$, corresponding to $\delta=0.05$ meV for $\gamma=0.5$meV.. 

We would like to note that a frequency stabilized laser may be used as a reference for highly stable pump. 
In this case, for example, a fractional frequency stability $\sim5\times10^{-16}/\sqrt{\tau_m}$ ($\tau_m$ is measurement time in seconds) has been demonstrated in Ref.~\cite{jiang2011making}. 
For polaritons, a typical laser energy is $\sim2$ eV, which gives a fractional frequency variation $2.5\times10^{-5}$ in our case. 
In the case of highly stable lasers, for $\tau_m=0.1$ ps, the fractional frequency stability is $\sim1.6\times10^{-9}$, which is much smaller than what our system requires.

\begin{figure}[h]
\includegraphics[width=0.95\columnwidth]{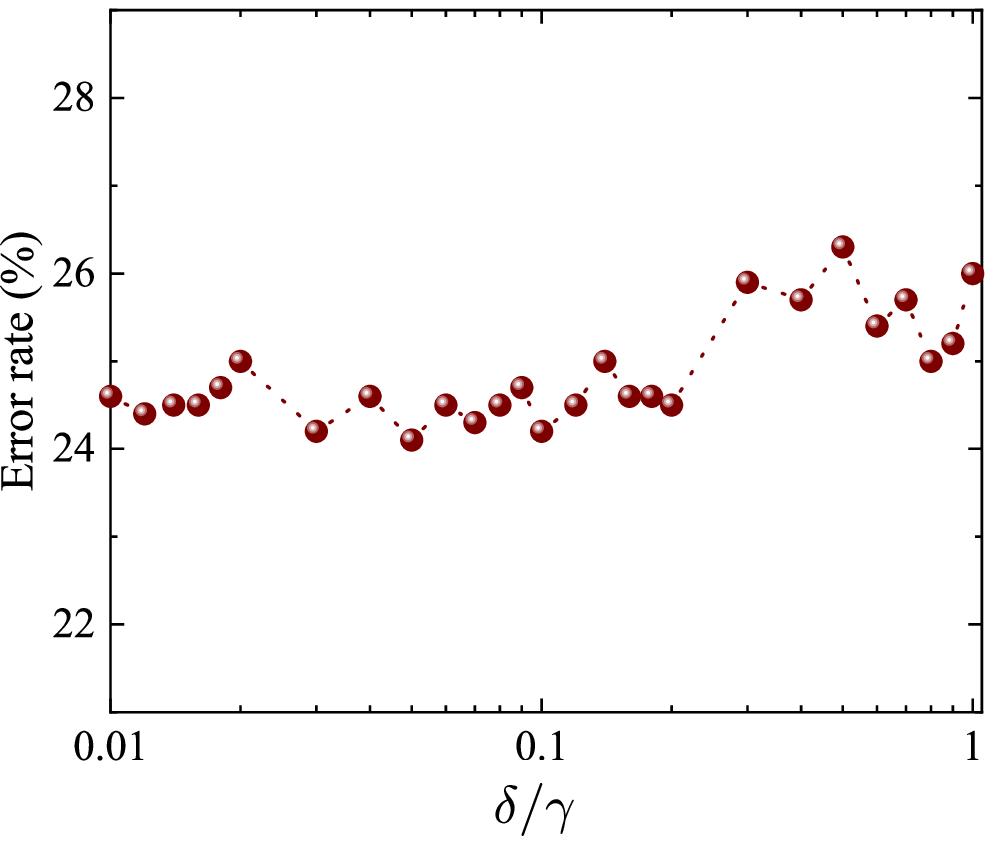}
\caption{Dependence of classification error rate on the fluctuation strength of the driving frequency for the single photon pumping case.}
\label{FigureB3}
\end{figure}

The two photon pumping can be realized with an inverse four-wave mixing process, e.g., see Ref.~\cite{kyriienko2014triggered}. 
This theoretically proposed configuration is similar to routinely studied four-wave mixing in microcavities~\cite{Saba2001,Wu2021,Baumberg2000,Sanvitto2006,Savvidis2000,Kundermann2003}, only instead of considering a polariton amplifier with simultaneous laser excitation of signal and pump modes, one considers simultaneous excitation of signal and idler modes, which results in effective two particle excitation of the pump mode.
In such case $P=\beta\sqrt{n_s n_i}$, where $\beta$ is the polariton-polariton interaction strength, $n_s$ is the density of polaritons in the signal state, and $n_i$ is the density in the idler state. 
As $\sqrt{n_s n_i}$ is effectively a measure of the density in both signal and idler states (i.e., the geometric mean), a typical value of $P$ would correspond to the typical blueshifts measured in experiments under resonant excitation. 
Indeed these can be anywhere between 0 and 1 meV. 
Varying this value is equivalent to changing the strength of laser fields, which is routinely done, potentially with an error as small as $0.02\%$, see for example Ref.~\cite{abbaspour2015effect}. 
In our case, the smallest change of P is $\sim0.2$ meV, which is much higher than $2\times10^{-4}$  meV mentioned above.

Last, we also note that the total time should be within the polariton lifetime. Our simulations indicate that if we go beyond the polariton lifetime, the error rate increases. This is intuitive as if we go beyond the polariton lifetime, the information from the earlier input signals fed into the system is already lost before the later signals are implemented. This is a key difference from a classical system, where an incoherent pump could compensate the loss due to finite lifetime. In a quantum reservoir, this is not possible as incoherently injected polaritons will not have the same quantum correlation with rest of the polaritons injected by the input signal.

\renewcommand{\thefigure}{C\arabic{figure}}
\begin{figure*}
\setcounter{figure}{0}
\includegraphics[width=1.9\columnwidth]{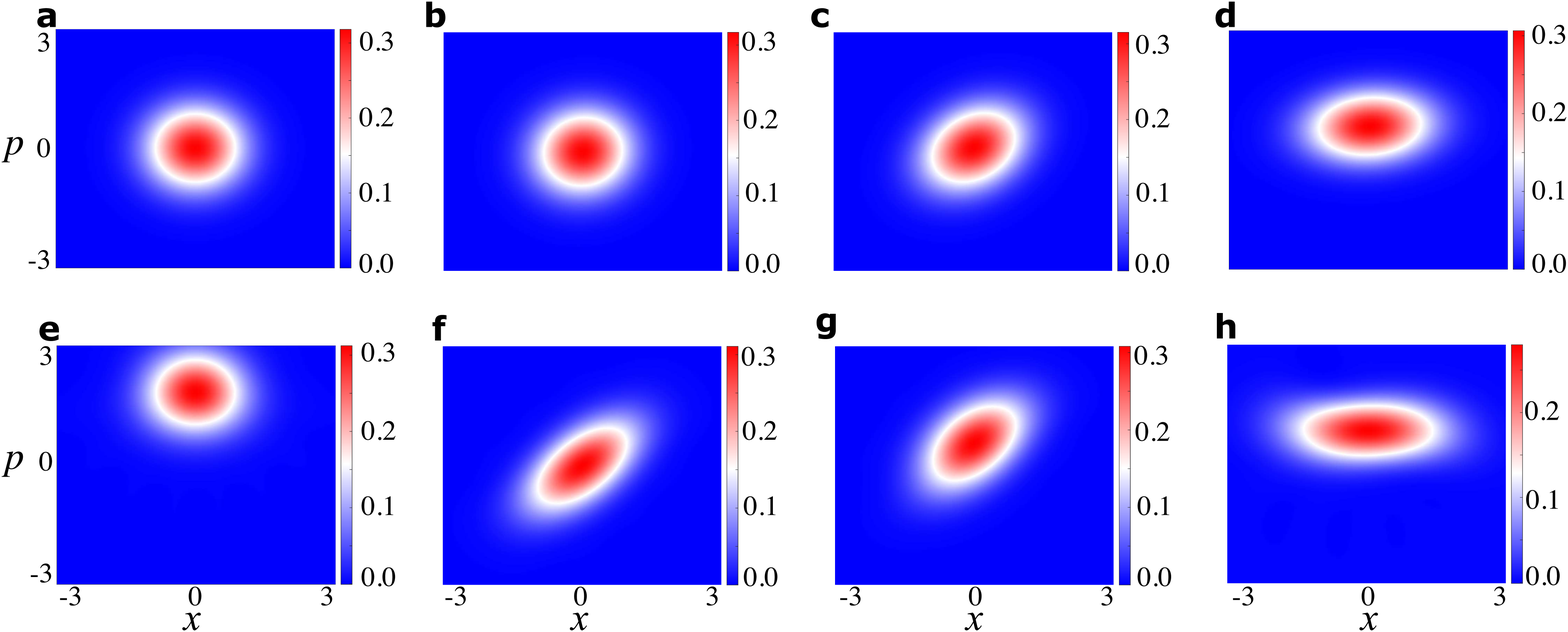}
\caption{Distribution of Wigner function under different pumping methods. 
Panel \textbf{a}-\textbf{d} correspond to the Wigner function with $\mathcal{SP}$, $\mathcal{TP}$, $\mathcal{SPTP}$ and $\mathcal{SPTP}^{\prime}$ pumping methods, respectively. 
Panels \textbf{e}-\textbf{h} are the corresponding ones with higher pumping strength.}
\label{FigureC1}
\end{figure*}

\section{APPENDIX C: Wigner Function of Different Pumping Methods}
As shown in the main text, we have a clear improvement by utilising different pumping methods for the quantum reservoir network, the origin of which we now analyse. 
As we use Wigner function as the output from the quantum reservoir network, we shall check its distribution for each of the pumping schemes and strength. 
Fig. \ref{FigureC1}\textbf{a}, \textbf{b}, \textbf{c} and \textbf{d} show the quasiprobability distribution in phase space under weaker pumping strength for the scenario of $\mathcal{SP}$, $\mathcal{TP}$, $\mathcal{SPTP}$, and $\mathcal{SPTP}^{\prime}$, respectively (see Eq.~(\ref{Equation2}) in the main text and the explanation that follows). 
Furthermore, by increasing the pumping strength, the corresponding Wigner functions appear as in Fig.~\ref{FigureC1}\textbf{e}-\textbf{h}.

It is clear from the $\mathcal{SP}$ pumping scheme in Fig.~\ref{FigureC1} panels \textbf{a} and \textbf{e}, that by increasing the pumping strength, the Wigner function retains the Gaussian-shape, but the center moves.
In this case, for different input signals, the center of quasiprobability distribution in phase space is driven to different positions. 
A different mechanism happens for the $\mathcal{TP}$ pumping method, see Fig.~\ref{FigureC1}\textbf{b} and \textbf{f}. 
For different pumping strength, the distribution stays in the center but they have different ``squeezed'' shapes.
Hence, for $\mathcal{SP}$ and $\mathcal{TP}$ one can think that, in the phase space, we have one degree of freedom for each method, which is position ($\mathcal{SP}$) and squeezing ($\mathcal{TP}$). 
To some extent, this explains why the error rates of $\mathcal{SP}$ and $\mathcal{TP}$ pumping methods are quite close.

However, if we look at Fig.~\ref{FigureC1}\textbf{c} and \textbf{g}, with stronger pumping strength, the position and the squeezing of the distribution both change, which means that under the combined pumping schemes $\mathcal{SPTP}$, we have two degrees of freedom. 
With this scheme, different input signals will provide us with more varied outputs, and this improves the classification process. 
We go further to the fourth pumping method ($\mathcal{SPTP}^{\prime}$), in which we also encode the input information with a phase $\Theta$. 
As shown in Fig.~\ref{FigureC1}\textbf{d} and \textbf{h}, when we increase the pumping strength, we obtained a Wigner function with different center position, squeezing strength and direction. With these three degrees of freedom, we are able to get a classification error rate as low as $13\%$.

The Wigner functions shown in Fig.~\ref{FigureC1} are not negative (which is normally used to indicate the presence of quantumness).
We note that quantum enhancement in this paper is associated with the exploitation of the largely available Fock space as compared to the classical multi-network polariton computer.
The different pumping schemes then serve as different ``encoding'' methods in which information is fed into the largely available space in the quantum case.

\section*{Acknowledgement}
%\vspace{0.5cm}
{This work was supported by the Singapore Ministry of Education under its AcRF Tier 2 grant MOE2019-T2-1-004.}

\bibliography{references}
\end{document}